\documentstyle[12pt,twoside,fleqn,espcrc1]{article}

\newcommand{\AmS}{{\protect\the\textfont2
  A\kern-.1667em\lower.5ex\hbox{M}\kern-.125emS}}

\newcommand{\lsim}{\raisebox{-0.7ex}{$\stackrel{\textstyle <}{\sim}$ }}

\newcommand{\denn}{\mbox{$m_\Lambda - m_N$}}
\newcommand{\hf}{\mbox{$h_F$}}
\newcommand{\hd}{\mbox{$h_D$}}

\title{$NNK$ and $\Lambda\Lambda K$ Amplitudes in Chiral Perturbation Theory}

\author{M.J. Savage\address{Department of Physics and Astronomy,  Box 351560, \\
        University of Washington, Seattle, WA 98195-1560, USA}%
        \thanks{Talk presented at HYP97, Brookhaven National Laboratory,
        October 12-18, 1997.
        Partially supported by the Department of Energy Grant DE-FG03-97ER41014}
        and 
        R.P Springer \address{Department of Physics, Duke University, \\
        Durham, NC 27708, USA}}

\begin{document}
\maketitle

\begin{abstract}
We explore the  S-wave and P-wave amplitudes for  $NNK$ and 
$\Lambda\Lambda K$
interactions using chiral perturbation theory 
to ${\cal O}(m_s\log m_s)$.
In contrast to the large corrections found in 
P-wave amplitudes in the nonleptonic decay of octet baryons,
the $NNK$ amplitudes 
receive only small corrections to their 
tree-level values.
The uncertainty in the $\Lambda\Lambda K$ amplitudes is 
found to be large.
\end{abstract}

\section{Introduction}

The motivation for studying the $NNK$ and $\Lambda\Lambda K$ amplitudes 
is two-fold.   
First, one would like to determine if the problems chiral perturbation 
theory ($\chi PT$) has in describing the P-wave amplitudes 
for the nonleptonic 
decay of the octet-baryons, $B\rightarrow B^\prime\pi$, 
at leading order represent a failure of 
$\chi PT$  or   just an accident\cite{ejnon}.  
Second, one wishes to accurately determine these amplitudes 
to better understand the decay of hypernuclei.

The  on-shell $NNK$ and $\Lambda\Lambda K$ interactions 
are not directly observable.
In order to make theoretical predictions for processes
involving these interactions the symmetries of QCD are used
to relate them to other observables
such as 
$B\rightarrow B^\prime\pi$.
$\chi PT$ provides a natural framework to  do this 
in the limit of exact SU(3) and also allows one to systematically
include SU(3) breaking effects arising from the mass difference between the 
strange and the up and down quarks.
Local counterterms are analytic functions of the 
light quark masses while loop graphs 
involving the lightest octet of pseudo-Goldstone bosons 
give rise to contributions
that are non-analytic in the light quark masses.
These non-analytic  contributions are unique and 
dominate   local counterterms in the  chiral limit.

$\chi PT$ does poorly in describing the P-wave 
amplitudes of $B\rightarrow B\pi$ at leading order.
Table 1   shows the experimental and theoretical amplitudes both at 
loop and tree-level.  
The nonleptonic weak couplings are fit to the experimentally measured 
S-wave amplitudes in $B\rightarrow B\pi$, and  together with the 
experimentally measured 
axial matrix elements give rise to  theoretical predictions for the 
P-wave amplitudes\cite{ejnon}.
\begin{table}[hbt]
\newlength{\digitwidth} \settowidth{\digitwidth}{\rm 0}
\catcode`?=\active \def?{\kern\digitwidth}
\caption{S-wave and P-wave amplitudes for the nonleptonic decay of octet baryons}
\label{tab:amps}
\begin{tabular*}{\textwidth}{@{}l@{\extracolsep{\fill}}rrrrrr}
\hline
                 & \multicolumn{3}{c}{S-wave} 
                 & \multicolumn{3}{c}{P-wave} \\
\cline{2-4} \cline{5-7}
                 & \multicolumn{1}{r}{expt} 
                 & \multicolumn{1}{r}{tree (fit)} 
                 & \multicolumn{1}{r}{loop (fit)} 
                 & \multicolumn{1}{r}{expt} 
                 & \multicolumn{1}{r}{tree} 
                 & \multicolumn{1}{r}{loop} \\
\hline
$\Sigma^+\rightarrow n\pi^+$     
&$0.06$	 
&$0.0$	
&$0.00\pm 0.02$ 
&$1.81$	 
&$-0.18$	
&$0.82$ 
 \\
$\Sigma^+\rightarrow p\pi^0$  	
&$-1.43$	 
&$-1.4\pm 0.1$	
&$-1.3\pm 0.1$ 
&$1.17$	 
&$-0.35$	
&$0.36$ 
\\
$\Sigma^-\rightarrow n\pi^-$  	
&$1.88$	 
&$2.0\pm 0.2$	
&$1.9\pm 0.2$ 
&$-0.06$	 
&$0.32$	
&$0.34$ 
\\  
$\Lambda\rightarrow p\pi^-$  	
&$1.43$	 
&$1.5\pm 0.2$	
&$1.4\pm 0.2$ 
&$0.52$	 
&$-0.56$	
&$-0.52$ 
\\  
$\Lambda\rightarrow n\pi^0$  	
&$-1.04$	 
&$-1.1\pm 0.1$	
&$-1.0\pm 0.1$ 
&$-0.39$	 
&$0.40$	
&$0.38$ 
\\  
$\Xi^-\rightarrow \Lambda\pi^-$  	
&$-1.98$	 
&$-1.9\pm 0.1$	
&$-2.0\pm 0.2$ 
&$0.48$	 
&$-0.13$	
&$0.35$ 
\\  
$\Xi^0\rightarrow \Lambda\pi^0$  	
&$1.52$	 
&$1.4\pm 0.1$	
&$1.4\pm 0.1$ 
&$-0.33$	 
&$0.18$	
&$-0.24$ 
\\  
\hline
\end{tabular*}
\end{table}
It was suggested in \cite{ejnon}\ that the failure of 
leading order $\chi PT$
to reproduce the P-wave amplitudes does not represent a failure of 
$\chi PT$ but rather is an accident.   
There are two pole graphs that contribute to the 
P-wave amplitudes at leading order and for the measured value 
of couplings there is 
substantial cancellation between them.
Understanding this cancellation is beyond  $\chi PT$, 
but it is suggestive that
it is accidental in this system.   
In addition, one sees that the non-analytic corrections 
${\cal O}(m_s \log m_s)$, bring the amplitudes into  better, 
but still poor agreement.
The uncertainties  in the theoretical predictions of the 
P-wave amplitudes
are currently under investigation by one of us\cite{rpsnon}.

\section{The $NNK$ Amplitudes}

There are three vertices that occur in $\Delta s=1$ weak 
nonleptonic 
interactions involving nucleons and kaons;
$p \bar p K^0$, $n \bar p K^+$, and $n \bar n K^0$.  
Isospin symmetry relates these amplitudes
\begin{eqnarray}
{\cal A}^{(L)}(nnK)  - {\cal A}^{(L)}(ppK) = {\cal A}^{(L)}(npK) 
\ \ \ ,
\end{eqnarray}
where $L=0$ (S-wave) or $L=1$ (P-wave).
The amplitudes can be written in terms of contributions at a given order
in perturbation theory
\begin{eqnarray}
{\cal A}^{(L)} = {\cal A}_0^{(L)} + {\cal A}_1^{(L)} + \cdots \ ,
\end{eqnarray}
where the subscript denotes the order in chiral perturbation 
theory and the 
dots denote higher orders.

The $\Delta s=1$ weak interactions of the pseudo-Goldstone bosons 
and  the lowest lying baryons are described, assuming octet
dominance, by the Lagrange density 
\begin{eqnarray}\label{weakl}
{\cal L} &=& 
G_Fm_\pi^2 f_\pi \Big( h_D {\rm Tr} {\overline B}_v 
\lbrace \xi^\dagger h\xi \, , B_v \rbrace \; 
+ \; 
 h_F {\rm Tr} {\overline B}_v  
{[\xi^\dagger h\xi \, , B_v ]} \; 
+  h_C {\overline T}^\mu_v
(\xi^\dagger h\xi) T_{v \mu}  \; \nonumber \\ &&
+ \; 
 { h_\pi \over 8} {\rm Tr} \left(  h \, \partial_\mu 
\Sigma 
\partial^\mu 
\Sigma^\dagger  \right) 
\ + \ \cdots\ \ \ \Big) \ \ \ \   ,
\end{eqnarray}
where
\begin{eqnarray}
h = \left(\matrix{0&0&0\cr 0&0&1\cr 0&0&0}\right)  \ \ \ ,
\end{eqnarray}
and the constants $f_\pi$, $h_D, h_F, h_\pi$ and $h_C$ are 
determined experimentally at a given order in the expansion.  
The octet baryon field is denoted by $B_v$ and the decuplet field by 
$T_v^\mu$\cite{man}.

\subsection{S-Wave Amplitudes}

At tree level the S-wave amplitudes 
appear directly from the 
first and second  terms in the 
weak Lagrangian of Eq.~\ref{weakl}\ ,  
\begin{eqnarray}\label{stree}
{\cal A}_0^{(S)} (p \bar p K^0) &=& h_F-h_D \ \ \ ,\ \ \ 
{\cal A}_0^{(S)} (p \bar n K^+) = h_F+h_D \ \ \ ,\ \ \ 
{\cal A}_0^{(S)} (n \bar n K^0) = 2 h_F 
\ \ \ \ .
\end{eqnarray}
The measured S-wave amplitudes for $B\rightarrow B^\prime\pi$
at tree level lead to  $h_D$ = --.58 and  
$h_F$ = 1.40\cite{ejnon}. 
Therefore, the tree level S-wave amplitudes for $NNK$ are
2.0, 0.8, and 2.8, respectively.
At one-loop the computation involves several loop graphs as discussed in 
\cite{savsprNNK} and the effects of the 
${\cal O}(m_s \log m_s)$ breaking are shown in Table 2.
One finds that the central value of each amplitude is 
suppressed compared to tree-level by $\sim 30\%$.   
The uncertainty for each amplitude is found to be relatively small.
The impact of these results has been explored in \cite{bpr}, and presented by 
A. Ramos at this workshop.
They find that the ratio of neutron induced to proton induced 
decays of $\Lambda$-hypernuclei and the total decay rate
are modified at the $10\%-15\%$ level by the reduced 
amplitudes shown in Table 2.

\subsection{P-Wave Amplitudes}

The tree level P-wave amplitudes come directly from 
pole graphs involving one weak vertex from Eq.~\ref{weakl} and 
one strong vertex
\begin{eqnarray}\label{ptree}
{{\cal A}_0^{(P)} (p \bar p K^0) \over \Lambda_\chi} &=& 
   -{(D-F)(h_D-h_F) 
\over m_N-m_\Sigma}
                 \nonumber \\
{{\cal A}_0^{(P)} (p \bar n K^+) \over \Lambda_\chi} &=& -{1 \over 
6}{(D+3F)(h_D+3h_F) 
                 \over m_N-m_\Lambda} + {1 \over 
2}{(D-F)(h_D-h_F) 
                 \over m_N-m_\Sigma}\nonumber \\
{{\cal A}_0^{(P)} (n \bar n K^0) \over \Lambda_\chi} &=& -{1 \over 
6}{(D+3F)(h_D+3h_F) 
                 \over m_N-m_\Lambda}-{1 \over 2}{(D-F)(h_D-h_F) 
                 \over m_N-m_\Sigma}
\ \ \ \ .
\end{eqnarray}

\noindent The numerical values for these amplitudes
are found by using tree-level parameters 
$h_D$ = --0.58, $h_F$ = 1.40, $D$ = 0.8, and $F$ = 0.5 \cite{ejnon,man}.  
This yields tree level P-amplitudes of  --2.4, 9.1, and 6.7 
$(\times \Lambda_\chi / 1 \, {\rm GeV})$, respectively.  
It is important to notice that there is only one pole graph contributing to 
each amplitude.
Therefore the cancellations that were found to be problematic in 
$B\rightarrow B^\prime \pi$ do not occur for the $NNK$ amplitudes.
These amplitudes are found to be of natural size in contrast to those for 
$B\rightarrow B^\prime \pi$.
Loop contributions to these amplitudes were computed to 
${\cal O}(m_s \log m_s)$ and found to be moderately small.
Measurement of the $NNK$ amplitudes  would provide valuable insight
into the applicability of $\chi PT$ for these processes.
If there was agreement then this would suggest that the problem with the 
P-wave amplitudes in $B\rightarrow B^\prime\pi$ is merely an 
accident and is not a failure of $\chi PT$.
If the agreement is poor then this suggests that we really do not have 
a good understanding of these processes.

\section{The $\Lambda\Lambda K$ Amplitudes}

The S-wave and P-wave amplitudes for $\Lambda\Lambda K$  are 
determined in the same way as the $NNK$ amplitudes\cite{savsprLL}.
At tree-level we have 
\begin{eqnarray}\label{streelam}
{\cal A}_0^{(S)} (\Lambda \Lambda K^0) & = & - h_D
\nonumber\\
{{\cal A}^{(P)}_0 (\Lambda \Lambda K^0) \over \Lambda_\chi}
 & =& {1 \over 6}{(D-3F)(3\hf\ -\hd\ ) \over m_\Lambda - m_\Xi \ }-
{1 \over 6}{(D+3F)(\hd+3\hf) \over \denn \ }
\ \ \ \ ,
\end{eqnarray}
and notice that there are contributions from two pole graphs as for
$B\rightarrow B^\prime \pi $.
The loop contributions of the form ${\cal O}(m_s \log m_s)$ 
are found to move the 
central value of the amplitudes only slightly from their tree-level 
values, but the uncertainties become very large.
It really only makes sense to quote a range for both the S-wave and 
P-wave amplitudes, as shown in Table 2.
In making predictions for the decay of doubly-$\Lambda$ hypernuclei 
these large uncertainties must be considered.

\begin{table}[hbt]
\caption{S-wave and P-wave amplitudes for $NNK$ and $\Lambda\Lambda K$ }
\label{tab:NNamps}
\begin{tabular*}{\textwidth}{@{}l@{\extracolsep{\fill}}rrrr}
\hline
                 & \multicolumn{2}{c}{S-wave} 
                 & \multicolumn{2}{c}{P-wave} \\
\cline{2-3} \cline{4-5}
                 & \multicolumn{1}{r}{tree} 
                 & \multicolumn{1}{r}{loop} 
                 & \multicolumn{1}{r}{tree} 
                 & \multicolumn{1}{r}{loop} \\
\hline
$nnK$     
&$2.8$	 
&$1.9\pm 0.4$	
&$6.7$	
&$6\pm 1$ 
 \\
$ppK$     
&$2.0$	 
&$1.5\pm 0.1$	
&$-2.4$	
&$-1.7\pm 0.2$ 
 \\
$npK$     
&$0.8$	 
&$0.4\pm 0.1$	
&$9.1$	
&$7\pm 1$ 
 \\
$\Lambda\Lambda K$     
&$0.6$	 
&$-0.5 \lsim {\cal A} \lsim 1.1$	
&$-5.2$	
&$-9 \lsim {\cal A} \lsim  -0.5$ 
 \\
\hline
\end{tabular*}
\end{table}

\section{Conclusions}

We have computed the ${\cal O}(m_s \log m_s)$ corrections to the 
S-wave and P-wave amplitudes for the $NNK$ and $\Lambda\Lambda K$ 
interaction.
The $NNK$ amplitudes are not sizably modified from their 
tree-level values, in contrast to the P-wave 
amplitudes for $B\rightarrow B^\prime \pi$.
Measurement of these amplitudes in hypernuclear decay  would provide a
valuable test of chiral perturbation theory.
The $\Lambda\Lambda K$ amplitudes are  poorly constrained at loop-level
and this large uncertainty must be included in predictions for 
the decay of doubly-$\Lambda$ hypernuclei.
The role of higher order counterterms, while not explicitly discussed here,
has been estimated in the theoretical uncertainty assigned in the 
fitting procedure, and  discussion can be found 
in \cite{savsprNNK,savsprLL}.

\bigskip\bigskip

This work was stimulated by several discussions  with Cornelius Bennhold.


\begin{thebibliography}{9}
\bibitem{ejnon} E. Jenkins, Nucl. Phys. {\bf B375} 561 (1992).


\bibitem{rpsnon} R. P. Springer, {\it work in progress} .

\bibitem{man}The formalism of heavy baryon chiral perturbation
theory is introduced in E. Jenkins and A.V. Manohar,
In ``Dobogokoe 1991, Proceedings, 
{\sl Effective field theories of the standard model}'' 113-137.

\bibitem{savsprNNK} M.J. Savage and R.P. Springer, Phys. Rev. C {\bf 53},
441 (1996).

\bibitem{bpr}  A. Parreno, A. Ramos, and C. Bennhold, 
Phys. Rev. C {\bf 56}, 339 (1997).


\bibitem{savsprLL} M.J. Savage and R.P Springer, 
to appear in Phys. Rev. {\bf C57} (1997).

\end{thebibliography}
\end{document}